\documentclass[a4paper,12pt]{article}

\usepackage{ifpdf}

\newif\ifpdf
\ifx\pdfoutput\undefined
  \pdffalse
\else
  \pdfoutput=1
  \pdftrue
\fi

\RequirePackage{xspace} %
\RequirePackage{subfigure} %
\RequirePackage[centertags]{amsmath} %
\RequirePackage{amssymb}
\RequirePackage{wrapfig} %
\RequirePackage{calc} %
\RequirePackage{ifthen}
\RequirePackage{tabularx} %
\RequirePackage{flafter} %
\RequirePackage{fancyhdr} %

\ifpdf
  \RequirePackage[pdftex]{color}%
  \RequirePackage{colortbl}%
  \RequirePackage{array}%
  \RequirePackage[pdftex]{graphicx}

  \RequirePackage[ pdftex, plainpages = false, pdfpagelabels,
                 pdfpagelayout = useoutlines,
                 bookmarks,
                 breaklinks = true,
                 linktocpage,
                 pagebackref,                      % to include page numbers in bibliography
                 colorlinks = true,
                 linkcolor = blue,
                 urlcolor  = blue,
                 citecolor = blue,
                 anchorcolor = blue,
                 hyperindex = true,
                 hyperfigures
                 ]{hyperref}

\else
  \RequirePackage{color}
  \RequirePackage{colortbl}
   \RequirePackage{array}
  \RequirePackage[dvips]{graphicx}
  \RequirePackage{hyperref}
  \usepackage{rotating}
\fi

\usepackage{makeidx} % enable indexing
\usepackage{setspace} %enable setting spacing, %e.g. singlespace, onehalfspace, and doublespace
\usepackage{rotating} %enable sidewaysfigure
\usepackage{ecltree}
\usepackage{epic}
\usepackage{supertabular}  % this will allow the table not to break
\usepackage{color}
\usepackage{exscale}
\usepackage{fontenc}
\usepackage{ifthen}
\usepackage{latexsym}
\usepackage{makeidx}
\usepackage{syntonly}
\usepackage{inputenc}
\usepackage{graphicx}
\usepackage{setspace}
\usepackage{caption2}
\usepackage[english]{babel}
\usepackage[square, comma,numbers,sort&compress]{natbib}
\usepackage{hypernat}
\usepackage{boxedminipage}
\usepackage{framed}
\usepackage{longtable}
\usepackage[all]{hypcap}    %included to make the hyperlink go to the top of figure
\usepackage{algorithm2e}
\usepackage{algorithmic}
\usepackage{lscape}
\usepackage{pdflscape}

\newcommand{\note}[1]{\marginpar[left]{\singlespace \tiny #1}}
\newcommand{\pois}{Poiseuille}
\newcommand{\Hs}      {\hspace{-0.5cm}} %
\newcommand{\CIF}     {\centering \includegraphics[width=2.5in]} %
\newcommand{\Vmin}    {\vspace{-0.2cm}} %

\renewcommand{\sectionmark}[1]%
      {\markright{\thesection\ #1}} %stops it capitalizing. #1 has value of section name

\renewcommand{\note}[1]{}

\doublespace %\onehalfspace

\title
{ %
\vspace*{3.0cm} \LARGE{\bf Comparing Poiseuille with 1D Navier-Stokes Flow in Rigid and Distensible Tubes and Networks} \vspace*{4.0cm} \\
}

\author{Taha Sochi\footnote{Imaging Sciences \& Biomedical Engineering, King's College London, The Rayne
Institute, St Thomas' Hospital, London, SE1 7EH, UK. Email: taha.sochi@kcl.ac.uk.} \vspace*{5.0cm}}

\begin{document}

\maketitle %
\pagenumbering{arabic}

\newpage
\phantomsection \addcontentsline{toc}{section}{Contents} %
\tableofcontents

%\newpage
%\phantomsection \addcontentsline{toc}{section}{List of Figures} %
%\listoffigures
%
%\phantomsection \addcontentsline{toc}{section}{List of Tables} %
%\listoftables

\newpage
\phantomsection \addcontentsline{toc}{section}{Abstract} \noindent
{\noindent \LARGE \bf Abstract} \vspace{0.5cm}\\
\noindent %

A comparison is made between the Hagen-Poiseuille flow in rigid tubes and networks on one side and
the time-independent one-dimensional Navier-Stokes flow in elastic tubes and networks on the other.
Analytical relations, a \pois\ network flow model and two finite element Navier-Stokes
one-dimensional flow models have been developed and used in this investigation. The comparison
highlights the differences between \pois\ and one-dimensional Navier-Stokes flow models which may
have been unjustifiably treated as equivalent in some studies.


%%%%%%%%%%%%%%%%%%%%%%%%%%%%%%%%%%%  Head style  %%%%%%%%%%%%%%%%%%%%%%%%%%%%%%%%%%%
\pagestyle{headings} %
\addtolength{\headheight}{+1.6pt}
\lhead[{Chapter \thechapter \thepage}]%
      {{\bfseries\rightmark}}
\rhead[{\bfseries\leftmark}]%
     {{\bfseries\thepage}} %tell it to put page number at rhead
\headsep = 1.0cm               % Added 07 Sep 2006
%%%%%%%%%%%%%%%%%%%%%%%%%%%%%%%%%%%%%%%%%%%%%%%%%%%%%%%%%%%%%%%%%%%%%%%%%%%%%%%%%%%%

\newpage
%XXXXXXXXXXXXXXXXXXXXXXXXXXXXXXXXXXXXXXXXXXXXXXXXXXXXXXXXXXXXXXXXX
\section{Introduction} \label{Introduction}

Poiseuille flow model has been widely used in the earth science and engineering studies for
modeling and simulating the flow of Newtonian fluids in networks of rigid tubes which may represent
a network of interconnected pipes for oil transportation or a simplified imitation of porous media.
The model is naturally extended to include Poiseuille-like flow of time-independent non-Newtonian
fluids \cite{Skellandbook1967, Sorbiebook1991, Valvatnethesis2004, Sochithesis2007, SochiB2008,
SochiArticle2010, SochiComp2010}. Although the 1D Navier-Stokes flow model for elastic tubes and
networks is the more popular \cite{FormaggiaGNQ2001, SmithPH2002, RuanCZC2003, SherwinFPP2003,
UrquizaBLVF2003, PontrelliR2003, MilisicQ2004, FernandezMQ2005, FormaggiaLTV2006, FormaggiaMN2006}
and normally is the more appropriate one for biological hemodynamic modeling, \pois\ model has also
been used in some studies for modeling and simulating blood flow in large vessels without
accounting for the distensibility of the biological networks.

One of the main differences between \pois\ and 1D flow models for single tube, which also affects
the network flow since the individual tubes in the network are subject to the same flow principles
as the stand-alone tubes, is that the flow in the first model depends on the pressure difference
while in the second model it depends on the actual pressure at the inlet and outlet
\cite{SochiElasticFlowTube2013}. This difference is mainly based on the rigidity and distensibility
of the flow ducts in these two types of tubes and networks. Another principal difference between
\pois\ and 1D models for network flow, which reflects their complexity and practical relevance and
originates from their single tube models, is that \pois\ is linear and hence it is numerically
solved in a single iteration, while the 1D model is nonlinear and hence it requires an iterative
process which may cause convergence instabilities leading to compromises associated with
considerable numerical errors and approximations.

Apart from the appropriateness of one of these models or the other for a given physical situation
(a reason that dictates which model must be used in a specific situation) there are certain
practical advantages and disadvantages of \pois\ and 1D network flow models. In general \pois\ is
easier to implement, moreover it incurs a relatively low computational cost, normally of the order
of $N^2$ of memory space where $N$ is the number of network nodes, while the 1D is more difficult
to implement with a high computational cost of the order of $4N^2$ of memory space. This memory
cost, associated with the previously-indicated solver iteration requirement, have obvious
consequences on the speed of operation and overall performance. The high computational cost of the
1D model in terms of memory space and CPU time can substantially increase by the demand of fine
meshing and the use of higher order interpolation.

On the other hand, the 1D model gives more detailed picture as it depicts the flux and pressure
fields over the spatial domain inside the tubes; opposite to the \pois\ model which can only
provide the average flow rate in the tubes and the pressure at the junctions. However, the pressure
at the interior points of the tubes can be obtained directly for \pois\ flow due to the linearity
of the pressure field; a feature that cannot be replicated in the 1D flow due to the nonlinearity
of the pressure field. Another advantage of the 1D Navier-Stokes model is that it can simulate
transient flow as well as steady state flow, while the \pois\ model is basically time-independent
and hence any time-dependent feature cannot be directly and dynamically replicated. However,
consideration of transient effects in \pois\ model may be imitated indirectly by the generation of
a sequence of time-independent flow frames which represent snapshots of the overall time-dependent
process.

%XXXXXXXXXXXXXXXXXXXXXXXXXXXXXXXXXXXXXXXXXXXXXXXXXXXXXXXXXXXXXXXXX
\section{\pois\ Model for Single Tube and Network} \label{PoisModel}

\pois\ formula for the flow of laminar, incompressible, axi-symmetric, time-independent,
fully-developed flow of Newtonian fluids in rigid cylindrical tubes assuming no-slip-at-wall
\cite{SochiSlip2011} conditions is given by

\begin{equation}\label{poisEq}
    Q=\frac{\pi r^4 \Delta P}{8\mu L}
\end{equation}
where $Q$ is the volumetric flow rate, $r$ is the tube radius, $\Delta P$ is the pressure drop
along the tube, $\mu$ is the fluid dynamic viscosity and $L$ is the tube length. This equation can
be derived by several methods \cite{Skellandbook1967, Sochithesis2007, SochiB2008,
SochiVariational2013}; most of which are based on the use of Navier-Stokes equation or one of its
subsidiaries.

There are several methods for modeling and implementing \pois\ flow in a network of rigid tubes.
The methods are mainly based on imposing a \pois\ flow condition defined by a constant conductance,
as given by Equation~\ref{poisEq}, on each tube in the network associated with a conservation of
flow condition on all junctions of the network, as well as boundary conditions on all the boundary
nodes. In the following we briefly describe one of these methods which may be the most
straightforward one and is based on imposing pressure boundary conditions.

For a network with $N$ nodes, which include all the internal junctions as well as the boundary
nodes, the following system of simultaneous linear equations is formed

\begin{equation}\label{PoisMatrixEq}
    \mathbf{C} \mathbf{P} = \mathbf{Q}
\end{equation}
where $\mathbf{C}$ is the tubes conductance matrix with dimensions of $N\times N$, $\mathbf{P}$ is
the pressure column vector with dimensions of $N\times 1$ and $\mathbf{Q}$ is primarily the total
flow column vector with dimensions of $N\times 1$. The rows of $\mathbf{C}$ basically represent the
relations describing the net volumetric flow rate on the nodes in the network, which, apart from
the boundary nodes, sums up to zero due to the incompressibility of fluid with the absence of
sources and sinks inside, while the columns represent the nodes of the network. Similarly, the rows
of the $\mathbf{P}$ vector represent the pressure at the $N$ nodes while the rows of the
$\mathbf{Q}$ vector primarily represent the net flow at the nodes. The pressure drop across each
tube is then split into two parts corresponding to the two nodes involved at the two ends of that
tube. All the entries in each row of the conductance matrix are set to zero except the ones related
to the tubes that are connected to the node represented by that row. By a proper choice of flow
signs, the total flow at each internal junction will be added to zero to satisfy the continuity
condition and hence the only sources and sinks for the flow are the boundary nodes. For the rows
corresponding to the boundary nodes, the boundary pressure conditions are imposed by setting the
diagonal entry of the conductance matrix to unity while all the other entries are set to zero with
the corresponding entry of the $Q$ vector being set to the given pressure boundary value of that
node.

\pois\ model for the networks can be validated by testing the satisfaction of \pois\ formula
(Equation~\ref{poisEq}) on all the tubes of the network given the obtained pressure at the two end
nodes of each tube, plus the satisfaction of the boundary conditions on the boundary nodes and the
continuity of flow at each junction node which requires that the net flow in each internal node
should be added to zero. A consequence of this is that the outflow (sum of volumetric flow rate at
outlet boundary nodes) should be equal to the inflow (sum of volumetric flow rate at inlet boundary
nodes).

%XXXXXXXXXXXXXXXXXXXXXXXXXXXXXXXXXXXXXXXXXXXXXXXXXXXXXXXXXXXXXXXXX
\section{1D Model for Single Tube and Network} \label{OnedModel}

The one-dimensional model for the time-independent flow of Newtonian fluids in elastic tubes is
derived from the following Navier-Stokes system, which is based on the mass and momentum
conservation principles

\begin{equation}
\frac{\partial Q}{\partial z}=0\,\,\,\,\,\,\,\,\,\,\,\,\, z\in\left[0,L\right]
\end{equation}

\begin{equation}
\frac{\partial}{\partial z}\left(\frac{\alpha Q^{2}}{A}\right)+\frac{A}{\rho}\frac{\partial
p}{\partial z}+\kappa\frac{Q}{A}=0\,\,\,\,\,\,\,\,\,\,\,\,\, z\in\left[0,L\right]
\end{equation}

In these equations, $A$ is the tube cross sectional area, $Q$ is the volumetric flow rate, $z$ is
the axial coordinate along the tube, $L$ is the length of tube, $\alpha$ ($=\frac{\int
u^{2}dA}{A\overline{u}^{2}}$ with $u$ and $\overline{u}$ being the fluid local and mean axial speed
respectively) is the correction factor for momentum flux, $\rho$ is the fluid mass density, $p$ is
the local pressure, and $\kappa$ is the viscosity friction coefficient usually given by $\kappa =
2\pi\alpha\nu/(\alpha-1)$ with $\nu$ being the fluid kinematic viscosity defined as the ratio of
the dynamic viscosity $\mu$ to the mass density. The model is based on the assumption of a laminar
fully-developed flow of incompressible fluids with a tube having a pressure-dependent cross
sectional area.

The relation between the pressure and flow rate in elastic tubes is dependent on the constitutive
relation that links the local pressure at a certain point along the tube axis to the cross
sectional area of the tube at that point. In reference \cite{SochiElasticFlowTube2013} the
following two flow relations have been derived based on two different pressure-area constitutive
relations

\begin{equation}\label{QElastic1}
Q=\frac{L-\sqrt{L^{2}-4\frac{\alpha}{\kappa}\ln\left(A_{ou}/A_{in}\right)\frac{\gamma}{3\kappa\rho}\left(A_{in}^{3}-A_{ou}^{3}\right)}}{2\frac{\alpha}{\kappa}\ln\left(A_{ou}/A_{in}\right)}
\end{equation}
and

\begin{equation}\label{QElastic2}
Q=\frac{-\kappa L+\sqrt{\kappa^{2}L^{2}-4\alpha\ln\left(A_{in}/A_{ou}\right)\frac{\beta}{5\rho
A_{o}}\left(A_{ou}^{5/2}-A_{in}^{5/2}\right)}}{2\alpha\ln\left(A_{in}/A_{ou}\right)}
\end{equation}

In these equations, $A_o$ is the tube unstretched cross sectional area at reference pressure, while
$A_{in}$ and $A_{ou}$ are the cross sectional area at the inlet and outlet respectively. Equations
\ref{QElastic1} and \ref{QElastic2} are based on the following pressure-area constitutive relations
respectively

\begin{equation}
p=\gamma\left(A-A_{o}\right)\label{pAEq1}
\end{equation}
and

\begin{equation}
p=\frac{\beta}{A_{o}}\left(\sqrt{A}-\sqrt{A_{o}}\right)
\end{equation}
where, in the last two equations, $p$ is the actual pressure relative to the reference pressure
with which the unstretched area in defined, and $\gamma$ and $\beta$ are the proportionality
coefficients that control the stiffness of the tube.

On multiplying the mass and momentum conservation equations by weight functions and integrating
over the solution domain the weak form of the Navier-Stokes flow system can be obtained. This weak
form, with suitable boundary conditions, can then be used as a basis for finite element
implementation in conjunction with an iterative scheme such as Newton-Raphson method. A detailed
account about the finite element formulation is given in references \cite{SochiTechnical1D2013,
SochiElasticFlowTube2013}.

The system can also be extended to a network of elastic tubes by imposing suitable pressure or flux
boundary conditions on all the boundary nodes, and compatibility and matching conditions on all the
internal junctions, where the latter conditions are derived from Riemann's method of
characteristics, and mass and energy conservation \cite{FormaggiaLQ2003, FormaggiaLTV2006,
SochiTechnical1D2013}.

The time-independent finite element solution of the 1D model on a network is validated by
fulfilling the boundary and matching conditions plus the time-independent analytical solution, as
given by Equations \ref{QElastic1} and \ref{QElastic2}, on each individual tube. The validation of
the time-independent finite element solution of the 1D model on a single tube is trivial as the
numeric solution can be compared directly to the analytic solution with the inspection of the
boundary conditions.

%XXXXXXXXXXXXXXXXXXXXXXXX
\section{Comparing \pois\ and 1D Models for Single Tube}

\begin{figure}
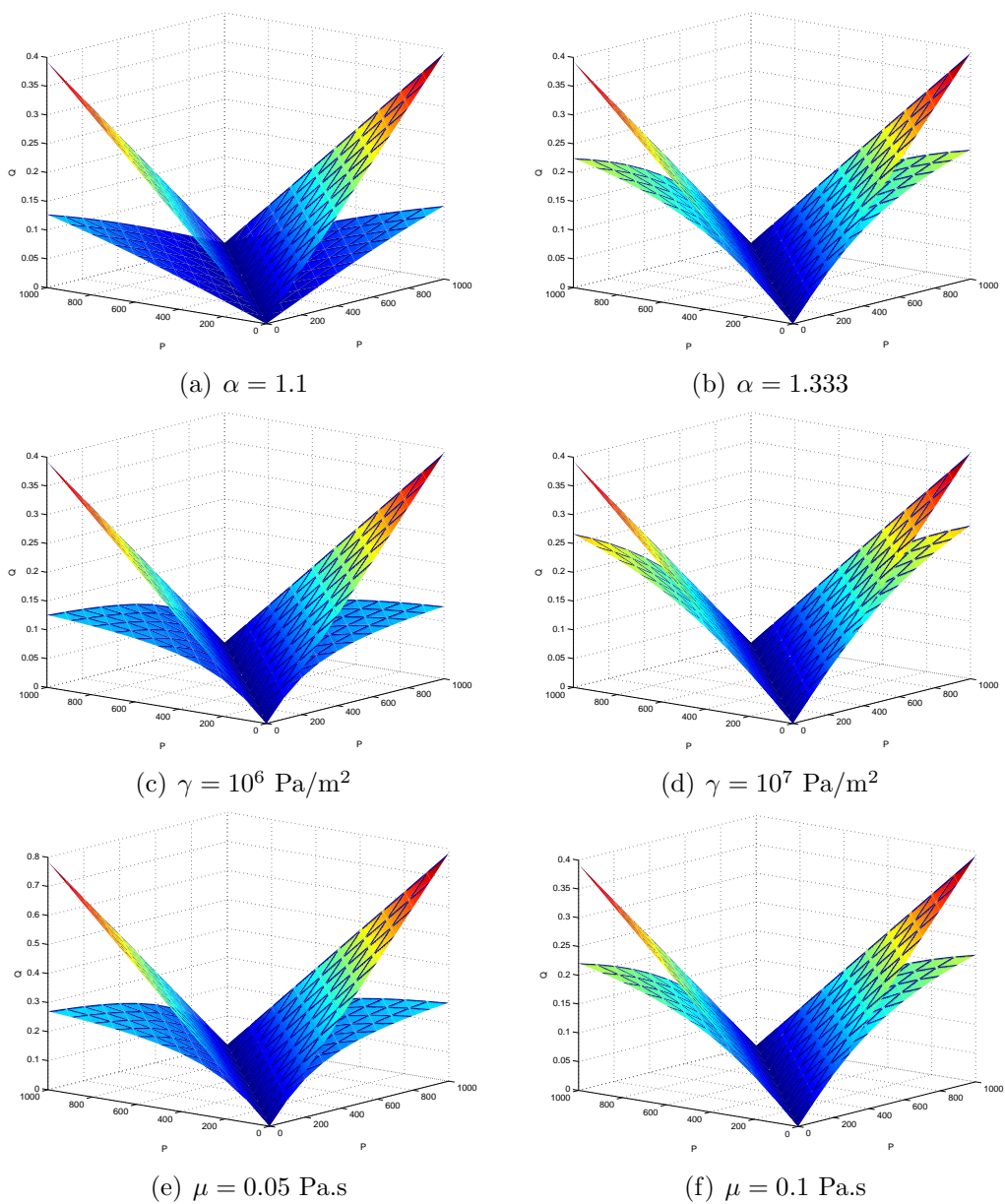

\centering %
\subfigure[$\alpha=1.1$]%
{\begin{minipage}[b]{0.5\textwidth} \CIF {alpha1}
\end{minipage}}
\Hs %
\subfigure[$\alpha=1.333$]%
{\begin{minipage}[b]{0.5\textwidth} \CIF {alpha2}
\end{minipage}} \Vmin

%XXXXXXXXXXXX
%
\centering %
\subfigure[$\gamma=10^6$~Pa/m$^2$]%
{\begin{minipage}[b]{0.5\textwidth} \CIF {gamma1}
\end{minipage}}
\Hs %
\subfigure[$\gamma=10^7$~Pa/m$^2$]%
{\begin{minipage}[b]{0.5\textwidth} \CIF {gamma2}
\end{minipage}} \Vmin

%XXXXXXXXXXXX
%
\centering %
\subfigure[$\mu=0.05$~Pa.s]%
{\begin{minipage}[b]{0.5\textwidth} \CIF {mu1}
\end{minipage}}
\Hs %
\subfigure[$\mu=0.1$~Pa.s]%
{\begin{minipage}[b]{0.5\textwidth} \CIF {mu2}
\end{minipage}}
\caption{The effect of $\alpha$, $\gamma$ and $\mu$ on the difference between Poiseuille model for
rigid tube and the 1D Navier-Stokes model for elastic tube using the first $p$-$A$ relation. The
planar surface belongs to Poiseuille while the curved surface belongs to the 1D. The horizontal
axes represent the pressure at the two ends of the tube where the inlet is taken where the largest
pressure occurs and hence the two surfaces are symmetric with respect to the diagonal
zero-flow-rate line of equal pressures. The pressures are in Pa while the flow rates are in
m$^3$/s. \label{Compare1}}
\end{figure}

\begin{figure}
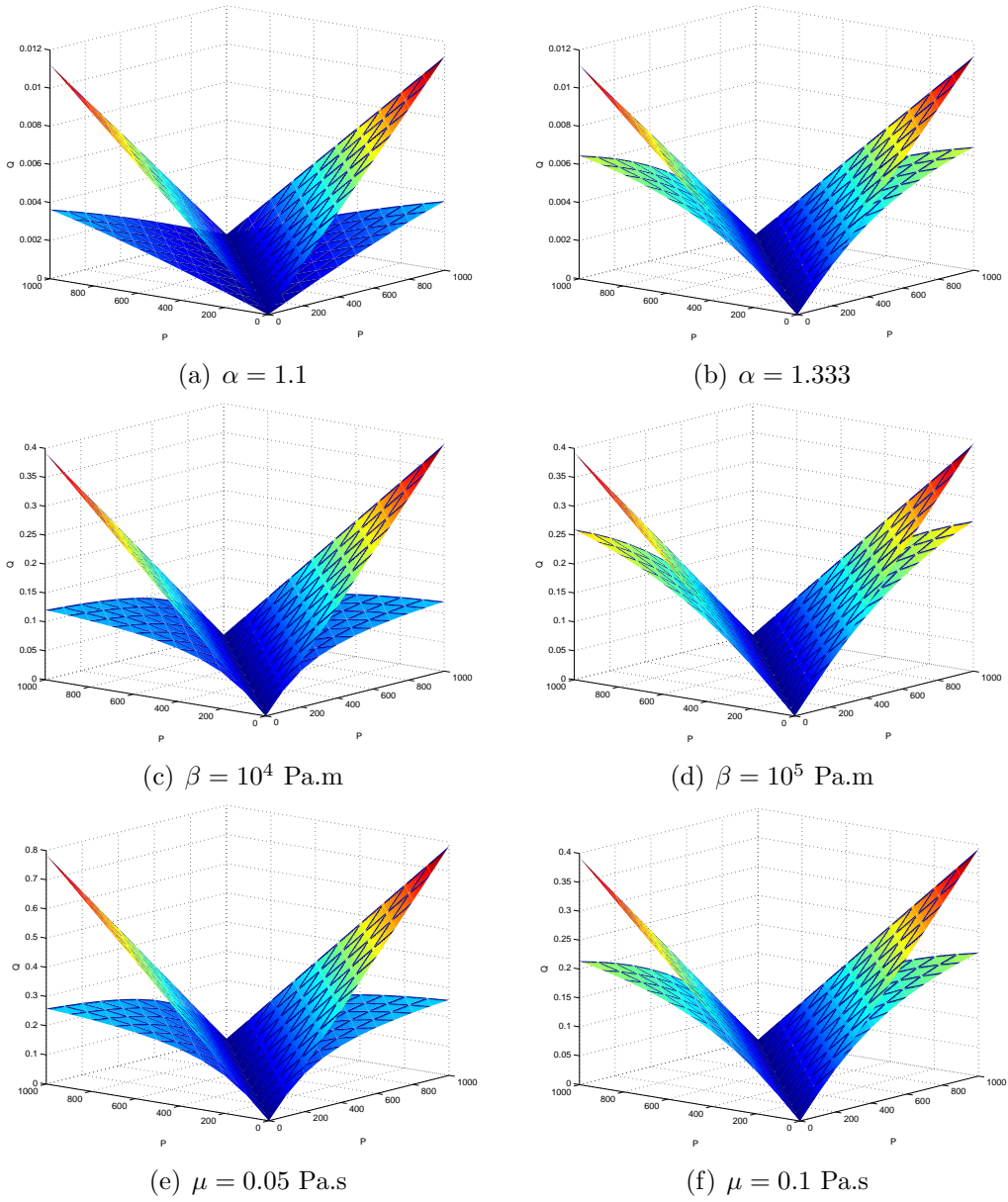

\centering %
\subfigure[$\alpha=1.1$]%
{\begin{minipage}[b]{0.5\textwidth} \CIF {alpha3}
\end{minipage}}
\Hs %
\subfigure[$\alpha=1.333$]%
{\begin{minipage}[b]{0.5\textwidth} \CIF {alpha4}
\end{minipage}} \Vmin

%XXXXXXXXXXXX
%
\centering %
\subfigure[$\beta=10^4$~Pa.m]%
{\begin{minipage}[b]{0.5\textwidth} \CIF {beta3}
\end{minipage}}
\Hs %
\subfigure[$\beta=10^5$~Pa.m]%
{\begin{minipage}[b]{0.5\textwidth} \CIF {beta4}
\end{minipage}} \Vmin

%XXXXXXXXXXXX
%
\centering %
\subfigure[$\mu=0.05$~Pa.s]%
{\begin{minipage}[b]{0.5\textwidth} \CIF {mu3}
\end{minipage}}
\Hs %
\subfigure[$\mu=0.1$~Pa.s]%
{\begin{minipage}[b]{0.5\textwidth} \CIF {mu4}
\end{minipage}}
\caption{The effect of $\alpha$, $\beta$ and $\mu$ on the difference between Poiseuille model for
rigid tube and the 1D Navier-Stokes model for elastic tube using the second $p$-$A$ relation.  All
the other aspects are as in Figure~\ref{Compare1}. \label{Compare3}}
\end{figure}

\begin{figure}
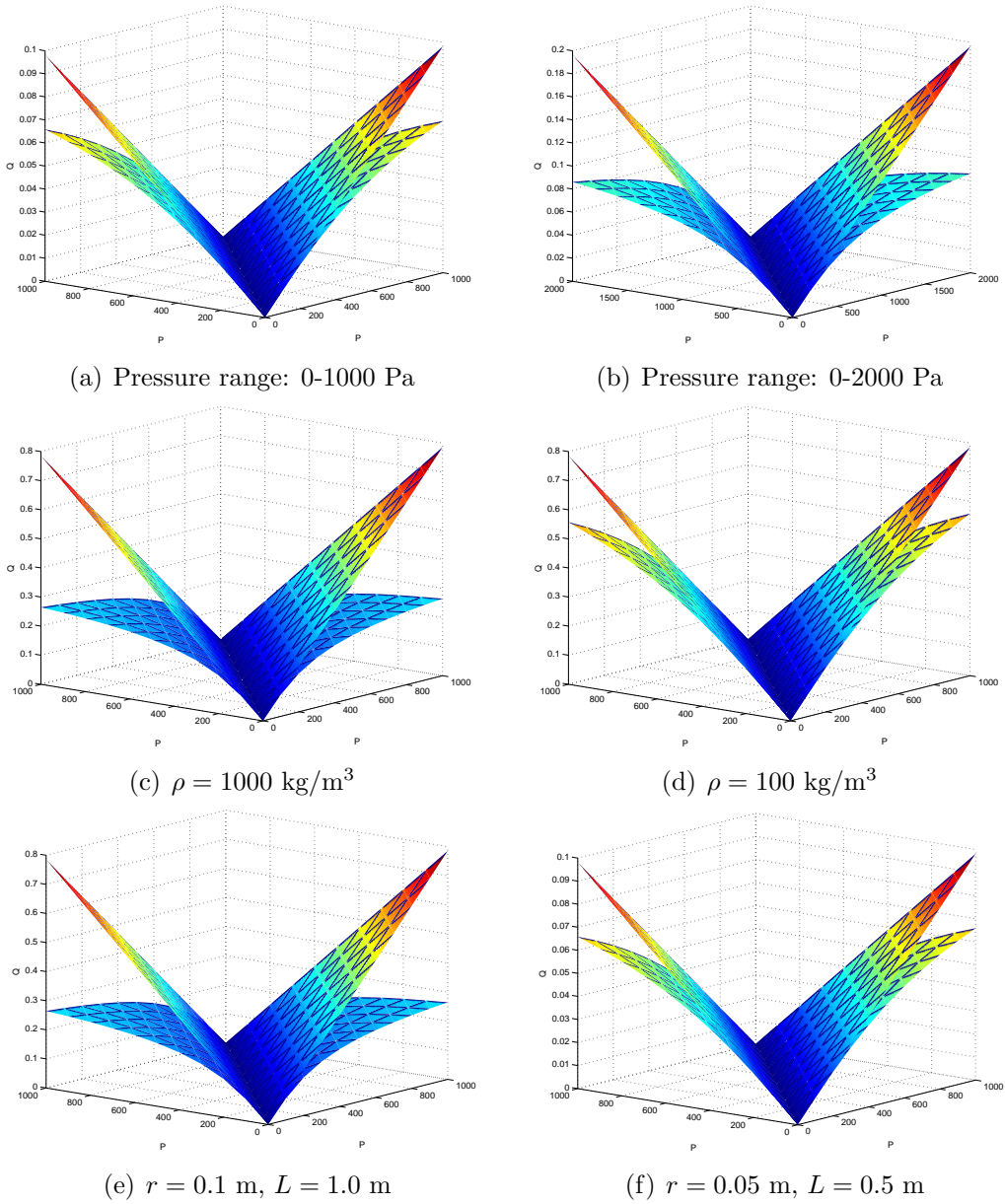

\centering %
\subfigure[Pressure range: 0-1000~Pa]%
{\begin{minipage}[b]{0.5\textwidth} \CIF {pres3}
\end{minipage}}
\Hs %
\subfigure[Pressure range: 0-2000~Pa]%
{\begin{minipage}[b]{0.5\textwidth} \CIF {pres4}
\end{minipage}} \Vmin

%XXXXXXXXXXXX
%
\centering %
\subfigure[$\rho=1000$~kg/m$^3$]%
{\begin{minipage}[b]{0.5\textwidth} \CIF {rho3}
\end{minipage}}
\Hs %
\subfigure[$\rho=100$~kg/m$^3$]%
{\begin{minipage}[b]{0.5\textwidth} \CIF {rho4}
\end{minipage}} \Vmin

%XXXXXXXXXXXX
%
\centering %
\subfigure[$r=0.1$~m, $L=1.0$~m]%
{\begin{minipage}[b]{0.5\textwidth} \CIF {size3}
\end{minipage}}
\Hs %
\subfigure[$r=0.05$~m, $L=0.5$~m]%
{\begin{minipage}[b]{0.5\textwidth} \CIF {size4}
\end{minipage}}
\caption{The effect of pressure range, $\rho$ and tube size on the difference between Poiseuille
model for rigid tube and the 1D Navier-Stokes model for elastic tube using the second $p$-$A$
relation. All the other aspects are as in Figure~\ref{Compare1}. \label{Compare4}}
\end{figure}

Thorough comparisons between rigid \pois\ and elastic 1D Navier-Stokes models have been carried out
for single tube as part of this study to investigate the effect of various parameters of these two
models on the flow behavior. The results of a sample of these comparisons, which represent the two
previously mentioned pressure-area constitutive relations, are plotted in Figures \ref{Compare1},
\ref{Compare2}, \ref{Compare3} and \ref{Compare4}. The \pois\ results shown in these figures are
obtained from the analytical solution (Equation~\ref{poisEq}) while the 1D results are obtained
from the analytical expressions (Equation~\ref{QElastic1} for the first $p$-$A$ relation and
Equation~\ref{QElastic2} for the second $p$-$A$ relation) with an endorsement by a finite element
numeric implementation \cite{SochiElasticFlowTube2013}.

As seen, the investigated flow, fluid and tube parameters; which include $\alpha$, $\beta$,
$\gamma$, $\mu$, pressure regime, $\rho$, and tube size; have significant effects on the flow
conduct of these two models and hence the results are highly dependent on these parameters. It is
noteworthy that all these reported results represent qualitative demonstration and hence may not
reflect a general trend due to the effect of other parameters which are held constant to
investigate the particular dependency. The variation of these parameters is expected to affect the
apparent trend in general. We also do not report the values of the other parameters for each one of
these cases in detail due to the generality of the current study and the qualitative nature of this
demonstration, as well as space limitation and avoiding unnecessary repetition. However, in all
these simulations, typical and representative values have been used for the parameters related to
the flow, fluid and flow paths, unless stated otherwise. This general conduct has also been
followed in the forthcoming investigation of the flow in networks.

%%%%%%%%%%%%%%%%%%%%%%%%%%%%%%%%%%%%%%%%%%%%%%%%%%%%%%%%%%%%%%%%%%%%%%%%%%%%%%%%%%%%%%%%%

%%%%%%%%%%%%%%%%%%%%%%%%%%%%%%%%%%%%%%%%%%%%%%%%%%%%%%%%%%%%%%%%%%%%%%%%%%%%%%%%%%%%%%%%%

\begin{figure}
\centering %
\subfigure[Pressure range: 0-1000~Pa]%
{\begin{minipage}[b]{0.5\textwidth} \CIF {pres1}
\end{minipage}}
\Hs %
\subfigure[Pressure range: 0-2000~Pa]%
{\begin{minipage}[b]{0.5\textwidth} \CIF {pres2}
\end{minipage}} \Vmin

%XXXXXXXXXXXX
%
\centering %
\subfigure[$\rho=1000$~kg/m$^3$]%
{\begin{minipage}[b]{0.5\textwidth} \CIF {rho1}
\end{minipage}}
\Hs %
\subfigure[$\rho=100$~kg/m$^3$]%
{\begin{minipage}[b]{0.5\textwidth} \CIF {rho2}
\end{minipage}} \Vmin

%XXXXXXXXXXXX
%
\centering %
\subfigure[$r=0.1$~m, $L=1.0$~m]%
{\begin{minipage}[b]{0.5\textwidth} \CIF {size1}
\end{minipage}}
\Hs %
\subfigure[$r=0.01$~m, $L=0.1$~m]%
{\begin{minipage}[b]{0.5\textwidth} \CIF {size2}
\end{minipage}}
\caption{The effect of pressure range, $\rho$ and tube size on the difference between Poiseuille
model for rigid tube and the 1D Navier-Stokes model for elastic tube using the first $p$-$A$
relation. All the other aspects are as in Figure~\ref{Compare1}. \label{Compare2}}
\end{figure}

%%%%%%%%%%%%%%%%%%%%%%%%%%%%%%%%%%%%%%%%%%%%%%%%%%%%%%%%%%%%%%%%%%%%%%%%%%%%%%%%%%%%%%%%%
%%%%%%%%%%%%%%%%%%%%%%%%%%%%%%%%%%%%%%%%%%%%%%%%%%%%%%%%%%%%%%%%%%%%%%%%%%%%%%%%%%%%%%%%%

%%%%%%%%%%%%%%%%%%%%%%%%%%%%%%%%%%%%%%%%%%%%%%%%%%%%%%%%%%%%%%%%%%%%%%%%%%%%%%%%%%%%%%%%%

%%%%%%%%%%%%%%%%%%%%%%%%%%%%%%%%%%%%%%%%%%%%%%%%%%%%%%%%%%%%%%%%%%%%%%%%%%%%%%%%%%%%%%%%%

\clearpage
%XXXXXXXXXXXXXXXXXXXXXXXX
\section{Comparing \pois\ and 1D Models for Network}

Both $p$-$A$ models have been implemented in a finite element network flow computer code as part of
this study. However, due to instabilities and convergence difficulties encountered in some cases of
the first model which may affect the reliability and generality of the results, we will only
present and analyze the results of the second model. This will not affect the generality of this
study since a single $p$-$A$ model is sufficient to highlight the main issues and draw the
conclusions; moreover the generic trends observed from both models are similar in general. In fact
the second model is the one which is widely used by the elastic network flow modelers, possibly due
to its numerical stability and reliability. The effect of the $p$-$A$ model on the \pois\ versus 1D
comparison may deserve a study by itself. It should be remarked that all the network results
reported in the current paper, like the single tube results, have been subjected to rigorous checks
using the validation criteria outlined in sections \ref{PoisModel} and \ref{OnedModel}.

\begin{figure}
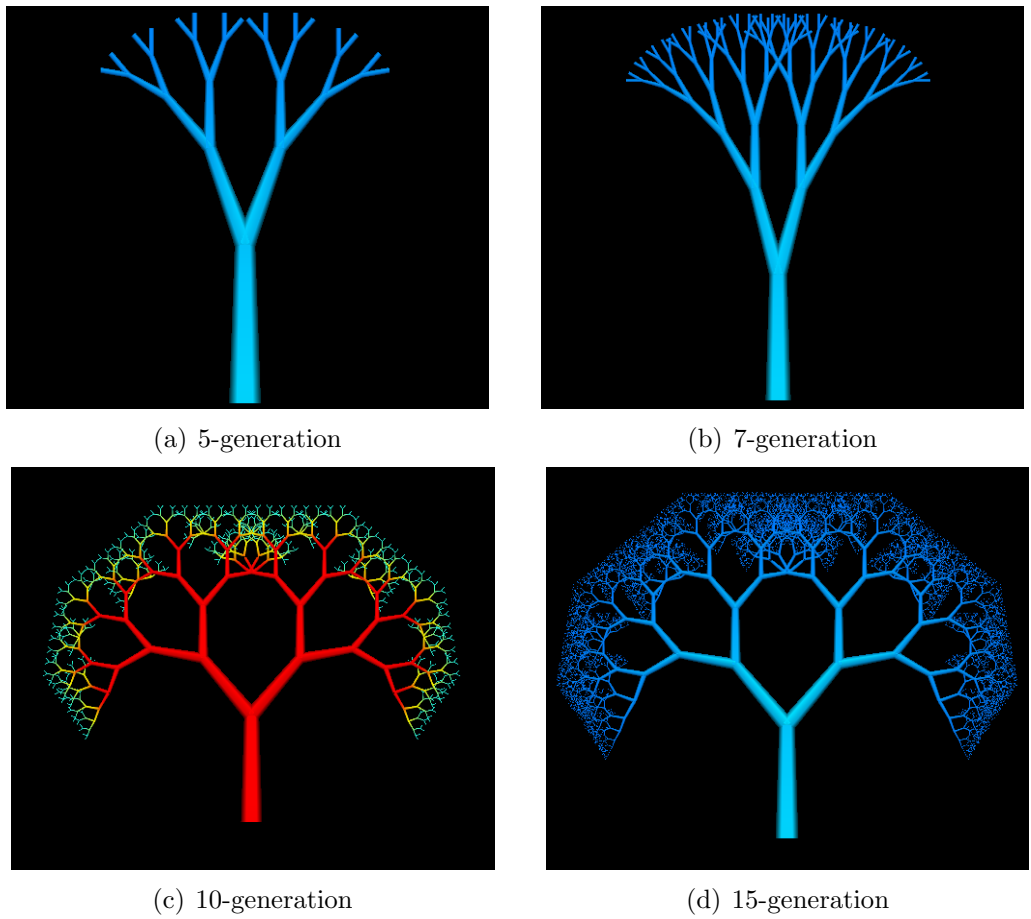

\centering %
\subfigure[5-generation]%
{\begin{minipage}[b]{0.5\textwidth} \CIF {Net5}
\end{minipage}}
\Hs %
\subfigure[7-generation]%
{\begin{minipage}[b]{0.5\textwidth} \CIF {Net7}
\end{minipage}} \Vmin

%XXXXXXXXXXXX
%
\centering %
\subfigure[10-generation]%
{\begin{minipage}[b]{0.5\textwidth} \CIF {Net10}
\end{minipage}}
\Hs %
\centering %
\subfigure[15-generation]%
{\begin{minipage}[b]{0.5\textwidth} \CIF {Net15}
\end{minipage}}
\caption{A sample of fractal networks used in the current investigation with different number of
generations as well as other branching parameters. \label{Networks}}
\end{figure}

With regard to the type of networks used in this investigation for \pois\ and 1D flow simulations,
we used fractal-type networks generated by a computer code. The main feature of these networks,
which differ in the number of generations and consequently the number of elements which ranges
between a few tubes to several thousands, is that they have a fixed branching angle with a constant
length to radius ratio; moreover the branching radius transition from one generation of tubes to
the next generation is subjected to a Murray-type rule \cite{SochiTechnical1D2013}, i.e.

\begin{equation}\label{Murray}
r_{m}^{\delta}=\sum_{i}^{n}r_{d_{i}}^{\delta}
\end{equation}
where $r_{m}$ is the radius of the mother tube, $r_{d_{i}}$ is the radius of the $i$th daughter
tube, $n$ is the number of daughter tubes, and $\delta$ is a constant index. Various networks with
different branching angle, length-to-radius ratio, Murray's index, and number of generations have
been used in our investigation. However, most of our networks were generated with $n=2$ and
$\delta$ between 2 and 3 with equal-size daughter tubes. A sample of these networks with different
number of generations are presented in Figure~\ref{Networks}. The reason for using highly regular
and symmetric fractal networks is that they, with their simplicity and regularity, reduce the
effect of factors related to the complex structure. With the use of these simple fractal networks
the flow results will essentially reflect the correlation of the flow with the varied parameters in
a simple manner. The use of networks with complex morphology will only obscure the results and
complicate the analysis due to the involvement of factors related to the complex geometry and
topology of the network.

\begin{figure}
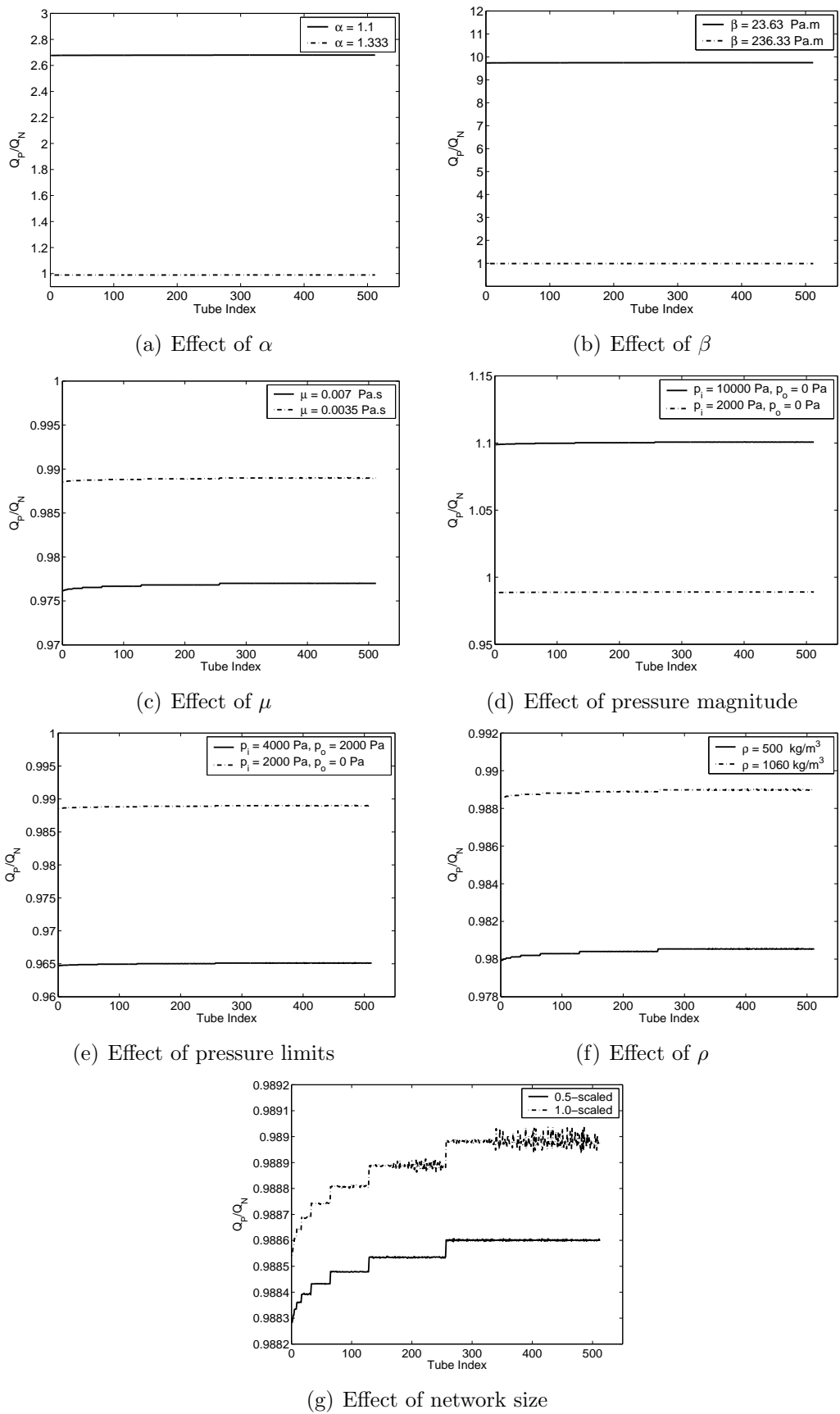

\vspace{-0.8cm}
\centering %
\subfigure[Effect of $\alpha$]%
{\begin{minipage}[b]{0.5\textwidth} \CIF {alpha3n}
\end{minipage}}
\Hs %
\subfigure[Effect of $\beta$]%
{\begin{minipage}[b]{0.5\textwidth} \CIF {beta3n}
\end{minipage}} \Vmin

%XXXXXXXXXXXX
%
\centering %
\subfigure[Effect of $\mu$]%
{\begin{minipage}[b]{0.5\textwidth} \CIF {mu3n}
\end{minipage}}
\Hs %
\centering %
\subfigure[Effect of pressure magnitude]%
{\begin{minipage}[b]{0.5\textwidth} \CIF {pres3n}
\end{minipage}} \Vmin

%XXXXXXXXXXXX
%
\centering %
\subfigure[Effect of pressure limits]%
{\begin{minipage}[b]{0.5\textwidth} \CIF {presLim3n} \label{presLim3n}
\end{minipage}}
\Hs %
\subfigure[Effect of $\rho$]%
{\begin{minipage}[b]{0.5\textwidth} \CIF {rho3n}
\end{minipage}} \Vmin

%XXXXXXXXXXXX
%
\centering %
\subfigure[Effect of network size]%
{\begin{minipage}[b]{0.5\textwidth} \CIF {size3n}
\end{minipage}}
\caption{Comparing the effect of various parameters on \pois\ and 1D Navier-Stokes network flow
models using the second $p$-$A$ constitutive relation. The $y$-axis represents the ratio of \pois\
flow rate, $Q_P$, to 1D flow rate, $Q_N$. \label{Compare7}}
\end{figure}

Using these fractal networks in conjunction with the second $p$-$A$ model for the 1D Navier-Stokes,
a number of \pois\ and 1D simulations have been conducted using typical network, fluid and flow
parameters. In these simulations a typical case in which \pois\ and 1D produce virtually identical
results with $\alpha=1.333$, $\beta=236$~Pa.m, $\mu=0.0035$~Pa.s, $p_i=2000$~Pa, $p_o=0$~Pa, and
$\rho=1060$~kg/m$^3$ has been used as a reference case. To investigate the effect of the parameters
of concern ($\alpha$, $\beta$, $\mu$, $p$, $\rho$, and size) a single parameter was varied at a
time from this reference case and a comparison between \pois\ and 1D was made to identify the
effect of that parameter. The method of comparison is based on computing the ratio of the
volumetric flow rate between \pois\ and 1D flow for each tube and plotting this ratio versus tube
indices, as seen in Figure~\ref{Compare7}. Although using the ratio of pressure at each junction
node is sensible for making the comparison it has not been used here due to the fact that the
pressure at each node for the 1D model is tube-dependent because each tube at a specific junction
has its own pressure due to the involvement of Bernoulli energy conservation principle as a
matching condition. As for the boundary conditions which were employed in these simulations, a
single inlet node belonging to the largest single tube was used to impose an inlet Dirichlet-type
pressure boundary condition while all the other boundaries at the other end, which belong to the
smallest tubes representing the last generation of the network, were subjected to zero-pressure
boundary conditions. However, there is one exceptional case where the outlet pressure was set to a
non-zero value to investigate the effect of pressure limits, as seen in Figure~\ref{presLim3n}.

On inspecting Figure~\ref{Compare7} it can be seen that while $\alpha$ and $\beta$ have a
significant effect on the 1D flow model as compared to \pois, the other parameters have either
moderate or negligible effect. However, this may not be true in general due to the limitation of
this study and the number of cases investigated. The use of networks with more complex morphology
is expected to introduce other sources of discrepancy between the two models and exacerbate the
difference between them.

%%%%%%%%%%%%%%%%%%%%%%%%%%%%%%%%%%%%%%%%%%%%%%%%%%%%%%%%%%%%%%%%%%%%%%%%%%%%%%%%%%%%%%%%%

%%%%%%%%%%%%%%%%%%%%%%%%%%%%%%%%%%%%%%%%%%%%%%%%%%%%%%%%%%%%%%%%%%%%%%%%%%%%%%%%%%%%%%%%%

%%%%%%%%%%%%%%%%%%%%%%%%%%%%%%%%%%%%%%%%%%%%%%%%%%%%%%%%%%%%%%%%%%%%%%%%%%%%%%%%%%%%%%%%%

\newpage
%XXXXXXXXXXXXXXXXXXXXXXXXXXXXXXXXXXXXXXXXXXXXXXXXXXXXXXXXXXXXXXXXX
\section{Conclusions} \label{Conclusions}

The purpose of this study is to compare \pois\ model for rigid tubes and networks with the
time-independent 1D Navier-Stokes model for elastic tubes and networks based on investigating the
effect of the parameters of these two models related to the flow, fluid and flow ducts. The main
conclusion of the current investigation is that \pois\ and 1D models could produce very different
results and hence they should not be used interchangeably as it may happened in some studies. The
use of one model or the other should be based on the merit of that model and its suitability to
capture the physical reality and similar objective considerations, not on convenience and pragmatic
factors. Moreover, the results of these models should be assessed relying on independent metrics
such as consistency and compliance with experimental observations.

Apart from the main theme of this investigation, it has been observed that for the investigated
cases of single tube, all the investigated parameters; which include $\alpha$, $\beta$, $\gamma$,
$\rho$, $\mu$, pressure range, and tube size; have significant effects. With regard to the
networks, limited in this study to those with fractal character, it has been observed that $\alpha$
and $\beta$ have the most sizeable impact. However, the impact of each one of the investigated
parameters can be affected by the involvement of other factors such as network topology and
geometry. Moreover, other factors are expected to play a significant role in networks with more
complex morphology.

Other effects related, for example, to the converging-diverging nature of the flow ducts and
non-Newtonian rheology \cite{SochiFeature2010, SochiPower2011, SochiNavier2013}, are not considered
in this study. These and all other factors should contribute to the complexity of the situation and
the departure of the results of \pois\ and \pois-like models for rigid tubes and networks from the
results of the 1D models for distensible tubes and networks.

The comparison presented in the current paper is very general and lacks thoroughness due to
specificity of purpose and space limitation. The effect of each one of the investigated parameters,
as well as many other aspects not touched in this study such as the effect of morphology of the
networks and their statistical distribution, deserve a study on its own.

\clearpage
%XXXXXXXXXXXXXXXXXXXXXXXXXXXXXXXXXXXXXXXXXXXXXXXXXXXXXXXXXXXXXXXXXXX
\phantomsection \addcontentsline{toc}{section}{Nomenclature} %
{\noindent \LARGE \bf Nomenclature} \vspace{0.5cm}

\begin{supertabular}{ll}
$\alpha$                &   correction factor for axial momentum flux \\
$\beta$                 &   stiffness parameter in the second pressure-area relation \\
$\gamma$                &   stiffness parameter in the first pressure-area relation \\
$\delta$                &   Murray's law index \\
$\kappa$                &   viscosity friction coefficient \\
$\mu$                   &   fluid dynamic viscosity \\
$\nu$                   &   fluid kinematic viscosity \\
$\rho$                  &   fluid mass density \\
%$\varsigma$             &   Poisson's ratio of vessel wall \\
%$\boldsymbol{\omega}$   &   vector of test functions in the weak form of finite element \\
%$\Omega$                &   solution domain \\
%$\Omega_e$              &   solution domain of an element \\
%$\partial \Omega$       &   boundary of the solution domain \\
\\
$A$                     &   tube cross sectional area \\
%$A_{BC}$                &   boundary condition for vessel cross sectional area \\
$A_{in}$                &   tube cross sectional area at inlet \\
$A_o$                   &   tube cross sectional area at reference pressure \\
$A_{ou}$                &   tube cross sectional area at outlet \\
%$\mathbf{B}$            &   matrix of force terms in the 1D Navier-Stokes equations \\
$\mathbf{C}$            &  conductance matrix \\
%$E$                     &   Young's modulus of vessel wall \\
%$f(A)$                  &   function in pressure-area relation \\
%$\mathbf{F}$            &   flux matrix in the 1D Navier-Stokes equations \\
%$h$                     &   length of element \\
%$h_o$                   &   vessel wall thickness at reference pressure \\
%$\mathbf{H}$            &   matrix of partial derivative of $\mathbf{F}$ with respect to $\mathbf{U}$ \\
%$\mathbf{J}$            &   Jacobian matrix \\
%$l^T_{1,2}$             &   left eigenvalues of $\mathbf{H}$ matrix (?) \\
$L$                     &   tube length \\
$n$                     &   number of daughter tubes in Murray's law \\
%$\mathfrak{N}$          &   norm of residual vector \\
$p$                     &   local pressure \\
$\mathbf{P}$            &  pressure column vector \\
$p_{i}$                 &   inlet pressure \\
$p_{o}$                 &   outlet pressure \\
$\Delta P$              &   pressure drop along the tube \\
%$p$                     &   order of interpolating polynomial \\
%$p_A$                   &   amplitude of sinusoidal input pressure signal \\
%$p_o$                   &   reference pressure (?) \\
%$\Delta p$              &   pressure step \\
%$q$                     &   dummy index for quadrature point \\
$Q$                     &   volumetric flow rate \\
$\mathbf{Q}$            &  total flow column vector \\
$Q_N$                    &   volumetric flow rate of 1D Navier-Stokes model \\
$Q_P$                    &   volumetric flow rate of \pois\ model \\
%$Q_{BC}$                &   boundary condition for volumetric flow rate \\
$r$                     &   tube radius \\
$r_{d}$                 &   radius of daughter tube \\
$r_{m}$                 &   radius of mother tube \\
%$\mathbf{R}$            &   weak form of residual vector \\
%$S_{a}$                 &   analytic solution \\
%$S_{n}$                 &   numeric solution \\
%$t$                     &   time \\
%$\Delta t$              &   time step \\
$u$                     &   local axial speed of fluid \\
$\overline{u}$          &   mean axial speed of fluid \\
%$\mathbf{U}$            &   vector of finite element variables \\
%$\mathbf{\Delta U}$     &   vector of change in $\mathbf{U}$ \\
$z$                     &   tube axial coordinate \\
%\\
%1D                      &   one-dimensional \\
%FE                      &   finite element \\
%NS                      &   Navier-Stokes \\
%TD                      &   time-dependent \\
%TI                      &   time-independent \\

\end{supertabular}

\newpage
\phantomsection \addcontentsline{toc}{section}{References} %
\bibliographystyle{unsrt}

\end{document}